\documentclass[conference]{IEEEtran}
\IEEEoverridecommandlockouts

\usepackage{mystyle}

\newcommand{\nt}{{n_{\mathsf{t}}}}
\newcommand{\nr}{{n_{\mathsf{r}}}}
\newcommand{\nc}{n_{\mathsf{c}}}

\newcommand{\opt}{{\star}}

\newcommand{\Bc}{\B{c}}

\newcommand{\Bs}{\B{s}}

\newcommand{\Bx}{\B{x}}
\newcommand{\By}{\B{y}}
\newcommand{\Bz}{\B{z}}

\newcommand{\BH}{\B{H}}

\newcommand{\BQ}{\B{Q}}

\newcommand{\Balpha}{\bm{\alpha}}

\newcommand{\BPhi}{\bm{\Phi}}
\newcommand{\BPsi}{\bm{\Psi}}

\newcommand{\mynull}{\mathbf{0}} 

\DeclareMathAlphabet{\mathbit}{OML}{cmr}{bx}{it}
\DeclareMathAlphabet{\mathgoth}{U}{ygoth}{m}{n}
\DeclareMathAlphabet{\mathfrak}{U}{yfrak}{m}{n}
\DeclareMathAlphabet{\mathswab}{U}{yswab}{m}{n}

\newcommand{\B}[1]{\mathbit{#1}}

\newtheorem{mythm}{Theorem}

\DeclareMathOperator{\Real}{\mathfrak{R}}

\DeclareMathOperator{\Exp}{\mathsf{E}}
\newcommand{\intd}{{\,\operatorname{d}}}

\DeclareMathOperator{\Transpose}{T}
\DeclareMathOperator{\Hermitian}{\dagger}
\newcommand{\Tr}{{\Transpose}}
\newcommand{\He}{{\Hermitian}}
\DeclareMathOperator{\cov}{cov}
\DeclareMathOperator{\var}{var}

\begin{document}

\title{A Simple Capacity Lower Bound\\for Communication with Superimposed Pilots}

\author{\IEEEauthorblockN{Adriano Pastore}
\IEEEauthorblockA{Centre Tecnol\`ogic de Telecomunicacions de Catalunya (CTTC/CERCA)\\
E-mail: adriano.pastore@cttc.cat}
\thanks{This work was supported by the Catalan Government under grant 2017 SGR 1479 and by the Spanish Ministery of Economy and Competitiveness through project TEC2014-59255-C3-1-R (ELISA).}
}

\maketitle


\begin{abstract}
We present a novel closed-form lower bound on the Gaussian-input mutual information for noncoherent communication (i.e., in which neither transmitter nor receiver are cognizant of the fading state) over a frequency-flat fading channel with additive noise. Our bound yields positive (non-trivial) values even in the most challenging case of zero-mean fast fading, a regime in which the conventional approach of orthogonal time-multiplexed pilots is unavailing and for which, to the best of the author's knowledge, no simple analytical bound is known. Its derivation relies on endowing the transmit signal with a non-zero mean, which can be interpreted as a pilot symbol that is additively superimposed onto the information-bearing Gaussian signal. The optimal fraction of transmit power that one should dedicate to this pilot is computed in closed form and shown to tend to one half at low SNR and to a limit above $2-\sqrt{2} \approx 0.586$ at high SNR. We further show how one can refine the bound for the general case of non-zero mean fading. Finally, we state an extension of our bound to the MIMO setting and apply it to compare superimposed vs.~orthogonal pilots on the SISO Rayleigh block fading channel.
\end{abstract}
\IEEEpeerreviewmaketitle

\section{Introduction}

Consider the discrete-time scalar communication channel with time index $k \in \mathbb{Z}$ governed by the system equation
\begin{equation}   \label{system_equation}
	Y_k = H_k S_k + Z_k
\end{equation}
wherein $S_k \in \mathbb{C}$ stands for the input signal, and the fading and noise processes $\{H_k\}$ and $\{Z_k\}$ are independent and identically distributed (i.i.d.)\ over time. We refer to this channel as {\em fast fading}. Furthermore, it is assumed that both the transmitter and the receiver are cognizant of the {\em distributions} of $H_k$ and $Z_k$, but not of their realizations. We are thus in total absence of channel-state information (CSI) at both ends of the link. The capacity scaling of this channel for the case of Rayleigh fading under peak and average-power constraints was first studied in~\cite{TaEl97}, where it was shown that the capacity grows double-logarithmically with the signal-to-noise ratio (SNR). Lapidoth and Moser~\cite{LaMo03} then extended this result to more general fading distributions using duality-based upper bounds on capacity. For the special case of Rayleigh fading, it was proved by Abou--Faycal, Trott and Shamai in~\cite{AbTrSh01} that the capacity-achieving distribution of the signal amplitude $|S_k|$ is finitely supported. This result was subsequently extended to Rician fading under fourth-moment signaling constraints~\cite{GuPoVe05}.

In channels such as the one under consideration, the fact that both fading and noise are i.i.d.\ over time outrules any kind of time-multiplexed pilot scheme for estimating the fading state, since any training observation will be immediately outdated. As a consequence, the popular achievable rate expressions based on a penalty factor incurred by time-orthogonal pilot symbols such as~\cite{AsKoGu11} collapse to zero in the fast-fading limit, although mutual information does not. Even those achievable rate expressions and capacity lower bounds that are {\em not} based on explicit training schemes tend to zero in the fast-fading limit, such as that proposed by Furrer and Dahlhaus~\cite{FuDa07} for block-fading channels, or that proposed by Deng and Haimovich~\cite{DeHa07} for stationary fading channels.

As a remedy, the receiver can exploit a known non-zero mean of the transmit signal (signal bias), which is sometimes referred to as a {\em superimposed} pilot symbol, or {\em overlay} pilot. In this setting, M\'edard's widely popular lower bound~\cite{Me00} on noncoherent fading channel capacity with a non-zero line-of-sight component (sometimes referred to as {\em worst-case noise} lower bound) equals zero.
In fact, M\'edard's bound matches capacity in the coherent setting (perfect CSI) but is totally loose in the noncoherent setting (no CSI) whenever the fading is zero-mean, and thus only of interest for moderately\footnote{The loosely defined notion of ``moderately imperfect CSI'' may be understood as the situation in which the ratio of the channel estimation error variance over the channel variance is substantially smaller than the reciprocal of the SNR, as also suggested in~\cite{LaSh02}.} imperfect CSI.

Even the nearest-neighbor decoding schemes with superimposed pilots proposed by Asyhari and ten Brink~\cite{Aste14,Aste17} fail to achieve any positive rate in the fast-fading limit, because the estimate of the time-$k$ fading gain $H_k$ is based exclusively on channel outputs $Y_{k'}$ at time instants $k' \neq k$ \cite[Eq.~(24)]{Aste14}, \cite[Eq.~(25)]{Aste17}.

In the present article we overcome this limitation by deriving a lower bound on the input-output mutual information of the fast-fading channel~\eqref{system_equation} where we impose a non-central Gaussian distribution on the input signal $S_k$. By a variation of M\'edard's proof of the worst-case noise lower bound~\cite{Me00}, we derive a new lower bound that satisfies the desired property.

%

\section{System model}   \label{sec:system_model}

Consider a pair $(S,Y) = (\bar{X}+X,Y) \in \mathbb{C}^2$ of complex random variables where $\bar{X} \in \mathbb{C}$ is a constant, $X \sim \mathcal{N}_\mathbb{C}(0,P)$ is complex-valued circularly-symmetric zero-mean with variance $\Exp\bigl[|X|^2\bigr]=P$, and where conditionally on $X = x$, the variable $Y$ is given by
\begin{equation}   \label{system_equation_2}
	Y = H(\bar{X}+x) + Z.
\end{equation}
Here, the complex-valued random variables $H$ and $Z$ are drawn such that $H$, $X$ and $Z$ are mutually independent. We assume that $H$ and $Z$ are both zero-mean and have variance $\Exp\bigl[|H|^2\bigr] = \Exp\bigl[|Z|^2\bigr] = 1$. Note that the additive noise $Z$ need not be Gaussian.

The variables $X$ and $Y$ model the information-bearing input and the output, respectively, of the point-to-point single-antenna communication system introduced in~\eqref{system_equation}, that employs i.i.d.\ Gaussian codebooks, with a frequency-flat fast-fading zero-mean channel $H$ and memoryless additive white noise $Z$. The constant signal bias $\bar{X}$ can be interpreted as an additively superimposed pilot symbol.

Assuming that the fading and noise process are stationary ergodic, the mutual information $I(S;Y) = I(X;Y)$ represents the supremum of communication rates for which the average probability of error of a maximum-likelihood decoder, averaged over the ensemble of i.i.d.\ Gaussian codebooks, tends to zero as the code's blocklength tends to infinity. Consequently, this Gaussian-input mutual information $I(X;Y)$ can be viewed as an achievable rate, or as a lower bound on the ergodic average-power constrained Shannon capacity $C(P)$, that is,
\begin{equation}   \label{capacity_lower_bound}
	C(P) \geq I(X;Y).
\end{equation}
Unfortunately, neither of the quantities in~\eqref{capacity_lower_bound} is analytically tractable, let alone expressible in closed form. In the next section, we shall derive a simple lower bound on $I(X;Y)$.

\section{Fading with no line of sight}

\subsection{Mutual information lower bound}

As mentioned previously, M\'edard's bound~\cite{Me00,LaSh02} for a channel like~\eqref{system_equation_2} is equal to zero, since $H$ is assumed to be zero-mean. Granting the {\em input signal} $S$ a non-zero mean (i.e., a superimposed pilot) is no remedy either to produce a non-trivial bound on $I(X;Y)$, unless one considers a {\em variation} of M\'edard's argument, which the following theorem presents and whose proof is provided underneath.
\begin{mythm}[Main result]   \label{thm:MI_lower_bound}
The mutual information between the variables $X$ and $Y$ as described in Section~\ref{sec:system_model} is lower-bounded as
\begin{equation}   \label{MI_lower_bound}
	I(X;Y) \geq \log\left(\frac{\var(|Y|^2)}{\var(|Y|^2) - P|\bar{X}|^2}\right)
\end{equation}
provided that $H$ and $Z$ have finite fourth moments.
\end{mythm}
\begin{IEEEproof}
Denoting as $\hat{X}(Y)$ some arbitrary function of $Y$, the mutual information $I(X;Y)$ can be bounded as follows:
\begin{IEEEeqnarray}{rCl}
	I(X;Y)
	&=& h(X) - h(X|Y) \\
	&=& \log(\pi e P) - h(X-\hat{X}(Y)|Y)   \label{MI_lower_bounding_1} \\
	&\geq& \log(\pi e P) - h(X-\hat{X}(Y))   \label{MI_lower_bounding_2} \\
	&\geq& \log(P) - \log(\var(X-\hat{X}(Y))).   \label{MI_lower_bounding_3}
\end{IEEEeqnarray}
The second equality \eqref{MI_lower_bounding_1} holds since $X \sim \mathcal{N}_\mathbb{C}(0,P)$ and because differential entropy is invariant to translations; the first inequality \eqref{MI_lower_bounding_2} holds because conditioning does not increase entropy; the last inequality \eqref{MI_lower_bounding_3} holds because for a fixed variance the Gaussian distribution maximizes differential entropy.
Let us expand the variance term as follows:
\begin{IEEEeqnarray*}{rCl}
	\IEEEeqnarraymulticol{3}{l}{
		\var(X-\hat{X}(Y))
	} \\
	\quad &=& \Exp\bigl[|X-\hat{X}(Y)|^2\bigr] - \bigl|\Exp[X-\hat{X}(Y)]\bigr|^2 \\
	&=& P - 2\Real\{\Exp[X^*\hat{X}(Y)]\} + \Exp[|\hat{X}(Y)|^2] - \bigl|\Exp[\hat{X}(Y)]\bigr|^2 \\
	&=& P + \var(\hat{X}(Y)) - 2\Real\{\Exp[X^*\hat{X}(Y)]\}. \IEEEeqnarraynumspace\IEEEyesnumber
\end{IEEEeqnarray*}
At this point it becomes clear that any useful estimator $\hat{X}(Y)$ will need to positively correlate with $X$, i.e., such that $\Real\{\Exp[X^*\hat{X}(Y)]\} > 0$, for otherwise we would obtain a (trivial) negative lower bound in \eqref{MI_lower_bounding_3}. Therefore, a linear estimator of the form $\hat{X}(Y) = \alpha Y$ with some constant $\alpha$, as chosen in the derivation of M\'edard's lower bound~\cite{Me00}, will not do in our case. Instead, we choose the estimator to be of the form $\hat{X}(Y) = \alpha|Y|^2$. The resulting error variance can be computed as
\begin{multline}   \label{input_error_variance}
	\var(X-\hat{X}(Y)) \\
	= P + |\alpha|^2\var(|Y|^2) - 2\Real\bigl\{\alpha\Exp\bigl[X^*|Y|^2\bigr]\bigl\}
\end{multline}
wherein the cross-correlation term is $\Exp\bigl[X^*|Y|^2\bigr] = P \bar{X}^*$. To obtain the latter equality, bear in mind that $\Exp\bigl[(X^*)^2\bigr] = \Exp[X^2] = 0$ since the phase of $X$ is uniformly distributed and independent of the magnitude $|X|$.

In order to obtain the sharpest bound, we choose $\alpha$ so as to minimize the variance~\eqref{input_error_variance}. The optimal phase of $\alpha$ should match the phase of $\bar{X}$, such that $\Real\{\alpha \bar{X}^*\} = |\alpha||\bar{X}|$. The optimal magnitude $|\alpha|$ should then minimize the quadratic function
\begin{equation}
	|\alpha| \mapsto P + |\alpha|^2 \var(|Y|^2) - 2 |\alpha| P |\bar{X}|
\end{equation}
the minimizer of which is $|\alpha| = {P|\bar{X}|}/{\var(|Y|^2)}$ and yields a minimum value
\begin{equation}   \label{minimum_of_quadratic_function}
	\min_{\alpha\in\mathbb{C}} \var(X-\alpha|Y|^2)
	= P\left(1-\frac{P|\bar{X}|^2}{\var(|Y|^2)}\right).
\end{equation}
Combining \eqref{minimum_of_quadratic_function} with \eqref{MI_lower_bounding_3}, we obtain the desired result.
\end{IEEEproof}

Note that the term $\var(|Y|^2)$ depends on fourth moments of the noise and fading distribution. If we denote the kurtosis of a random variable $A$ as
\begin{equation}
	\kappa_{A}
	= \frac{\Exp\bigl[|A-\Exp[A]|^4\bigr]}{\Exp\bigl[|A-\Exp[A]|^2\bigr]^2}
\end{equation}
then the kurtosis of $H$ and of $Z$ are respectively $\kappa_{H} = \Exp\bigl[|H|^4\bigr]$ and $\kappa_{Z} = \Exp\bigl[|Z|^4\bigr]$. With these, the variance of $|Y|^2$ can be evaluated as
\begin{multline}   \label{var_Y_square}
	\var(|Y|^2)
	= (\kappa_{H}-1) |\bar{X}|^4 + (2\kappa_{H}-1) P(2|\bar{X}|^2+P) \\
	{ } + 2 (|\bar{X}|^2+P) + \kappa_{Z} - 1.
\end{multline}
For the case of Rayleigh fading $H \sim \mathcal{N}_\mathbb{C}(0,1)$ and additive Gaussian noise $Z \sim \mathcal{N}_\mathbb{C}(0,1)$, we have $\kappa_{H}=\kappa_{Z}=2$.


\subsection{How much training power is needed?}

Assume that the transmitter is subjected to an average power constraint, meaning that the sum of constant pilot power $|\bar{X}|^2$ and data signal power $P$ cannot exceed a prescribed signal-to-noise ratio $\rho$:
\begin{equation}
	|\bar{X}|^2 + P \leq \rho.
\end{equation}
It is evident from~\eqref{MI_lower_bound} that the bound vanishes to zero if either $|\bar{X}|^2 = 0$ or $P=0$. Therefore, we wish to optimally balance the power split between the superimposed pilot and the data signal. For this purpose, we set $|\bar{X}|^2 = \nu\rho$ and $P=(1-\nu)\rho$ where $\nu \in [0;1]$ represents the power fraction allocated to the pilot.
The variance of $|Y|^2$ can now be expressed as
\begin{equation}
	\var(|Y|^2)
	= \eta(\rho) - \kappa_{H} \rho^2 \nu^2
\end{equation}
where $\eta(\rho) = (2\kappa_{H}-1)\rho^2 + 2\rho + \kappa_{Z}-1$. The bound~\eqref{MI_lower_bound} with optimized power allocation now reads as
\begin{equation}
	\max_{0 \leq \nu \leq 1} \log\left(1+\frac{\rho^2\nu(1-\nu)}{\rho^2(1-\kappa_{H})\nu^2 - \rho^2\nu + \eta(\rho)}\right).
\end{equation}
Solving this maximization yields an optimal pilot share
\begin{equation}
	\nu_{\opt}(\rho)
	= \frac{\eta(\rho)-\sqrt{\eta(\rho)(\eta(\rho)-\rho^2\kappa_{H})}}{\rho^2\kappa_{H}}.
\end{equation}
This optimal share has low and high-SNR limits
\begin{subequations}
\begin{align}
	\lim_{\rho \to 0} \nu_{\opt}(\rho)
	&= \frac{1}{2} \\
	\lim_{\rho \to \infty} \nu_{\opt}(\rho)
	&= 2-\tfrac{1}{\kappa_{H}}-\sqrt{\left(2-\tfrac{1}{\kappa_{H}}\right)\left(1-\tfrac{1}{\kappa_{H}}\right)}.   \label{nu_high_SNR}
\end{align}
\end{subequations}
The latter is a decreasing function of $\kappa_{H} > 1$ and thus lies within bounds
\begin{equation}
	1
	> \lim_{\rho \to \infty} \nu_{\opt}(\rho)
	> 2-\sqrt{2} \approx 0.586
\end{equation}
irrespective of the value of $\kappa_H$. Figure~\ref{fig:optimal_split} shows the plot of $\nu_{\opt}(\rho)$ for Rayleigh fading and Gaussian noise. Since $\kappa_{H}=2$ in this case, we have $\lim_{\rho\to\infty}\nu_{\opt}(\rho) = (3-\sqrt{3})/2 \approx 0.634$. Figure~\ref{fig:R_vs_SNR_plot} shows the resulting mutual information lower bound~\eqref{MI_lower_bound} as a function of the SNR. As one can see, the curve is bounded in the SNR, much like the mutual information $I(X;Y)$ itself, as studied in detail in~\cite{ChHaKoMa04}. Its high-SNR limit, for Rayleigh fading and optimized pilot share $\nu_\star(\rho)$, is equal to
\begin{equation}
	\log\left( 1 + \frac{(3-\sqrt{3})(\sqrt{3}-1)}{12 - (3-\sqrt{3})(5-\sqrt{3})} \right)
	\approx 0.11167
\end{equation}
nats, or equivalently, $0.16111$ bits.
\begin{center}
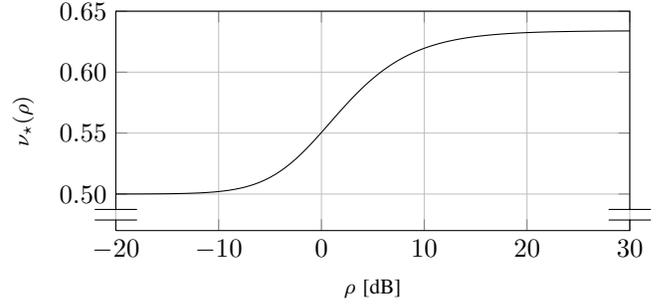
\begin{figure}[ht]
	\begin{tikzpicture}
\pgfplotsset{every axis/.append style={mark options=solid}}
\pgfplotsset{grid style={solid,ultra thin}}
\begin{axis}[
	grid=major,
	xmin=-20, xmax=30,
	ymax=0.65,
	height={4.5cm},
	width=.95\columnwidth,
	y tick label style={/pgf/number format/.cd,fixed,fixed zerofill,precision=2},
	ytick distance=0.05,
	xlabel={$\rho$ [dB]},
	ylabel={$\nu_{\opt}(\rho)$},
	axis y discontinuity=parallel,
	legend style={font=\footnotesize},
	label style={font=\footnotesize},
	enlarge y limits={value=0.2,lower},
]
	\pgfplotstableread{coord/coord_R.txt}\mytable
	\addplot[solid] table[x=SNRdB,y=nu_opt] {\mytable};
\end{axis}
\end{tikzpicture}
	\caption{Optimal fraction $\nu_{\opt}(\rho)$ of total transmit power $\rho$ allocated to the superimposed pilot for $H \sim \mathcal{N}_\mathbb{C}(0,1)$ and $Z \sim \mathcal{N}_\mathbb{C}(0,1)$.}
	\label{fig:optimal_split}
\end{figure}
\end{center}

\begin{center}
\begin{figure}[ht]
	\begin{tikzpicture}
\pgfplotsset{every axis/.append style={mark options=solid}}
\pgfplotsset{grid style={solid,ultra thin}}
\pgfplotsset{every axis y label/.append style={yshift=0pt}}
\pgfplotsset{every axis legend/.append style={cells={anchor=west},fill=white,at={(0.02,0.98)},anchor=north west}}
\begin{axis}[
	grid=major,
	xmin=-20,
	xmax=30,
	height={4.5cm},
	width=.95\columnwidth,
	xlabel={$\rho$ [dB]},
	ylabel={nats / channel use},
	y tick label style={/pgf/number format/.cd,fixed,fixed zerofill,precision=2},
	label style={font=\footnotesize},
	legend style={font=\footnotesize},
]
\pgfplotstableread{coord/coord_R.txt}\Rtable
\addplot[solid] table[x=SNRdB,y=I_opt] {\Rtable};
\addlegendentry{$\nu_{\opt}(\rho)$};
\addplot[dotted] table[x=SNRdB,y=I_subopt_low_SNR] {\Rtable};
\addlegendentry{$\nu=1/2$};
\addplot[dashed] table[x=SNRdB,y=I_subopt_high_SNR] {\Rtable};
\addlegendentry{$\nu=\frac{3-\sqrt{3}}{2}$};
\end{axis}
\end{tikzpicture}
	\caption{The bound from Theorem~\ref{thm:MI_lower_bound} against the SNR, plotted for three different choices of $\nu$.}
	\label{fig:R_vs_SNR_plot}
\end{figure}
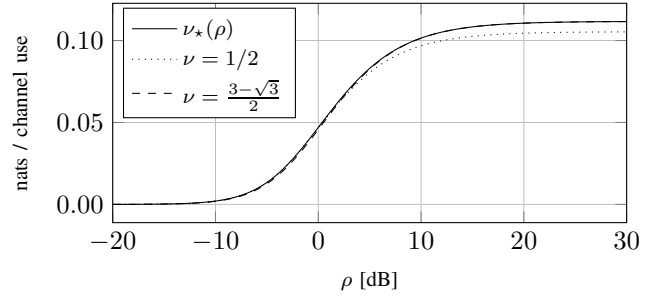
\end{center}

\section{Fading with a line of sight}   \label{sec:hybrid_capacity_lower_bound}

Consider the system equation~\eqref{system_equation_2}, but where the channel gain now includes a line-of-sight component in form of a non-zero mean $\bar{H}$ (while $H$ is still assumed zero-mean as before):
\begin{equation}   \label{system_equation_3}
	Y = (\bar{H}+H)(\bar{X}+x) + Z.
\end{equation}
It is worth mentioning that, while the mutual information bound from Theorem~\ref{thm:MI_lower_bound} was presented under the assumption of zero-mean fading (for it is the regime of most interest for that bound), the bound still holds for a non-zero-mean fading channel~\eqref{system_equation_3} as long as we replace the kurtosis $\kappa_H$ by the fourth moment $\Exp[|H|^4]$ throughout. However, for a system as in~\eqref{system_equation_3}, it might be more interesting to set $\bar{X}=0$ and use M\'edard's worst-case noise bound~\cite{Me00}
\begin{equation}   \label{Medard_bound}
	I(X;Y)
	\geq \log\left( 1 + \frac{ |\bar{H}|^2 P}{ \Exp[|H|^2] P + 1} \right)
\end{equation}
or its improved version~\cite{PaKoFo14}.
In summary, our bound~\eqref{MI_lower_bound} operates well in the regime $|\bar{H}|^2 \ll \Exp[|H|^2]$ and requires a superimposed pilot ($\bar{X} \neq 0$) whereas M\'edard's bound is useful in the opposite regime $|\bar{H}|^2 \gg \Exp[|H|^2]$ and requires no pilot ($\bar{X} = 0$). The derivation of the former relies on an input estimator of the form $\hat{X}(Y) = \alpha|Y|^2$ whereas M\'edard's bound was derived using $\hat{X}(Y) = \alpha Y$.

Can one construct a general bound that covers both regimes well and outperforms both the bound from Theorem~\ref{thm:MI_lower_bound} and M\'edard's bound? The solution is to choose an input estimator of the general form
\begin{equation}
	\hat{X}_{\Balpha}(Y)
	= \alpha_1 Y_1 + \alpha_2 Y_2
	= \Balpha^\Tr \By
\end{equation}
with complex coefficient vector $\Balpha^\Tr = \begin{bmatrix} \alpha_1 & \alpha_2 \end{bmatrix}$ and zero-mean observation vector
\begin{equation}
	\By
	= \begin{bmatrix} Y_1 \\ Y_2 \end{bmatrix}
	= \begin{bmatrix} |Y|^2 - \Exp[|Y|^2] \\ Y - \Exp[Y] \end{bmatrix}.
\end{equation}
Clearly, forcing either $\alpha_1$ or $\alpha_2$ to zero each recovers one of the two aforementioned bounds. A superior bound, however, is obtained when optimizing the pair $(\alpha_1,\alpha_2)$ as in
\begin{equation}
	I(X;Y)
	\geq \log(P) - \log\left( \min_{\Balpha} \var(X-\hat{X}_{\Balpha}(Y)) \right)
\end{equation}
in which the optimal coefficient vector $\Balpha_\opt$ is given by the Wiener-Hopf equation
\begin{equation}
	\Balpha_\opt^\Tr
	= \Exp[X \By^\He] \Exp[\By\By^\He]^{-1}
\end{equation}
and where, upon particularizing to the channel model~\eqref{system_equation_2}, we get
\begin{subequations}
\begin{IEEEeqnarray}{rCl}
	\Exp[X \By^\He]
	&=& \begin{bmatrix} (|\bar{H}|^2 + \Exp[|H|^2]) P \bar{X} & \bar{H}^* P \end{bmatrix} \\
	\Exp[\By\By^\He]
	&=&
	\begin{bmatrix}
		\var(|Y|^2)  & \Exp[Y^*|Y|^2] \\
		\Exp[Y|Y|^2] & \var(Y)
	\end{bmatrix}.   \label{Y_matrix}
\end{IEEEeqnarray}
\end{subequations}
Here, the entries of $\Exp[\By\By^\He]$ may all be evaluated in closed form. The resulting variance of $X-\hat{X}_{\Balpha_\opt}(Y)$ is
\begin{equation}
	\var(X-\hat{X}_{\Balpha_\opt}(Y))
	= P - \Exp[X\By^\He] \Exp[\By\By^\He]^{-1} \Exp[X^*\By]
\end{equation}
and the resulting capacity lower bound reads as
\begin{equation}
	I(X;Y) \geq
	\log\left( \frac{P}{P - \Exp[X\By^\He] \Exp[\By\By^\He]^{-1} \Exp[X^*\By]} \right).
\end{equation}
The latter can be straightforwardly generalized to any arbitrary vector $\By = \begin{bmatrix} Y_{1} & Y_{2} & \dots \end{bmatrix}^\Tr$ of functions $Y_{i} = f_{i}(Y)$ of the channel output, so we have the following general theorem:
\begin{mythm}   \label{thm:general_MI_lower_bound}
The mutual information between the variables $X$ and $Y$ as described in Section~\ref{sec:system_model} is lower-bounded as
\begin{equation}
	I(X;Y) \geq
	\log\left( \frac{P}{P - \Exp[X\By^\He] \cov(\By)^{-1} \Exp[X^*\By]} \right)
\end{equation}
provided that the covariance matrix $\cov(\By) \triangleq \Exp\bigl[(\By-\Exp[\By])(\By-\Exp[\By])^\He\bigr]$ is invertible. 
\end{mythm}

\begin{center}
\begin{figure}[ht]
	\begin{tikzpicture}
\pgfplotsset{every axis/.append style={mark options=solid}}
\pgfplotsset{grid style={solid,ultra thin}}
\pgfplotsset{every axis legend/.append style={cells={anchor=west},fill=white,at={(0.02,0.98)},anchor=north west}}
\begin{axis}[
	grid=major,
	xmin=0, xmax=1, ymin=0,
	height={4cm}, width={7cm},
	xlabel={$\lambda$},
	ylabel={[nats / channel use]},
	xtick distance={0.1}, ytick distance={0.1},
	xticklabels={,,,,,,,$0.6$,$0.7$,$0.8$,$0.9$,$1$},
	yticklabels={,,,,,,$0.5$,$0.6$,$0.7$},
	legend style={font=\footnotesize},
	label style={font=\footnotesize},
	scale only axis,
]
	\pgfplotstableread{coord/coord_I_hybrid.txt}\mytable
	\addplot[dashed,thick] table[x=alpha,y=I_Medard] {\mytable};
	\addlegendentry{M\'edard's bound~\eqref{Medard_bound}}
	\addplot[densely dotted,thick] table[x=alpha,y=I_simple] {\mytable};
	\addlegendentry{bound from Theorem~\ref{thm:MI_lower_bound}}
	\addplot[solid] table[x=alpha,y=I_hybrid] {\mytable};
	\addlegendentry{bound from Theorem~\ref{thm:general_MI_lower_bound}}
	\draw[fill=white,shift={(-4\pgflinewidth,-4\pgflinewidth)}] (0,0) rectangle (3.2cm,2.2cm);
\end{axis}
\begin{axis}[
	grid=major,
	xmin=0, xmax=0.15, ymin=0,
	height={2cm}, width={3cm},
	xticklabel style={/pgf/number format/.cd,fixed,fixed zerofill,precision=2,/tikz/.cd},
	yticklabel style={/pgf/number format/.cd,fixed,fixed zerofill,precision=2,/tikz/.cd},
	xticklabel={\ifdim\tick pt=0pt 0 \else\pgfmathprintnumber{\tick}\fi},
	yticklabel={\ifdim\tick pt=0pt 0 \else\pgfmathprintnumber{\tick}\fi},
	xtick distance={0.05},
	ytick distance={0.05},
	scale only axis,
	scaled y ticks=false,
]
	\pgfplotstableread{coord/coord_I_hybrid.txt}\mytable
	\addplot[dashed,thick] table[x=alpha,y=I_Medard] {\mytable};
	\addplot[densely dotted,thick] table[x=alpha,y=I_simple] {\mytable};
	\addplot[solid] table[x=alpha,y=I_hybrid] {\mytable};
\end{axis}
\end{tikzpicture}
	\caption{Comparison of bounds for a noncoherent Rician fading channel as a function of the Rician factor. The fading channel is modeled as consisting of a (squared) line-of-sight component $|\bar{H}|^2 = \lambda$ and an unknown Rayleigh-fading component of variance $\Exp[|H|^2] = 1-\lambda$. The SNR is set to $\rho = 1$ ($0$ dB). The plot region for low $\lambda$ is zoomed in for better visibility.}
	\label{fig:I_hybrid_plot}
\end{figure}
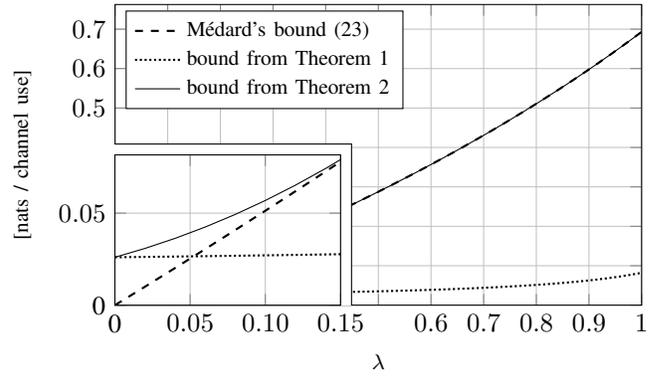
\end{center}

In Figure~\ref{fig:I_hybrid_plot}, M\'edard's bound is compared against the bound from Theorem~\ref{thm:MI_lower_bound} for a noncoherent Rician fading channel in which we let the ratio between the line-of-sight component and the (unknown) Rayleigh component vary with a parameter $\lambda \in [0;1]$. Accordingly, the cases $\lambda=0$ and $\lambda=1$ may be interpreted as the extreme situations of {\em no CSI} and {\em perfect CSI}, respectively.
We see that for a vanishing line-of-sight component ($\lambda=0$), M\'edard's bound vanishes, whereas our bound attains a positive value. The situation is reversed at the other end ($\lambda=1$), where M\'edard's bound becomes tight with perfect-CSI capacity ($\log(2) \approx 0.69$ nats) whereas our bound~\eqref{MI_lower_bound} is loose. The solid curve shows the bound from Theorem~\ref{thm:general_MI_lower_bound}, which outperforms both other bounds.

\section{Extension to the MIMO setting}

Theorem~\ref{thm:MI_lower_bound} can be generalized to the MIMO setting. Consider the input variable $\Bs = \bar{\Bx} + \Bx \in \mathbb{C}^{\nt}$ with $\bar{\Bx}$ a constant superimposed pilot and $\Bx \sim \mathcal{N}_\mathbb{C}(\mynull,\BQ)$, and where conditionally on $\Bx = \Bx'$, the output variable $\By \in \mathbb{C}^{\nr}$ is given by
\begin{equation}
	\By = \BH(\bar{\Bx} + \Bx') + \Bz
\end{equation}
with mutually independent $\BH$, $\Bx$ and $\Bz$.
\begin{mythm}[MIMO bound]   \label{thm:MI_lower_bound_MIMO}
The mutual information between $\Bx$ and $\By$ is lower-bounded as
\begin{equation}
	I(\Bx;\By)
	\geq \log|\BQ| - \log|\BQ - \BPhi\BPsi^{-1}\BPhi^\dagger|
\end{equation}
where $\BPhi = \Exp\bigl[ \Bx\Bc^\dagger\By\By^\dagger \bigr]$, $\BPsi = \cov(\By\By^\dagger\Bc)$ and $\Bc \in \mathbb{C}^{\nr}$ is an arbitrary constant vector chosen such that $\BPsi$ is invertible.
\end{mythm}
The proof is omitted but follows the same lines as that of Theorem~\ref{thm:MI_lower_bound}. Note that one can readily recover Theorem~\ref{thm:MI_lower_bound} by appropriately particularizing Theorem~\ref{thm:MI_lower_bound_MIMO}.

\subsection{An application: SISO Rayleigh block fading}

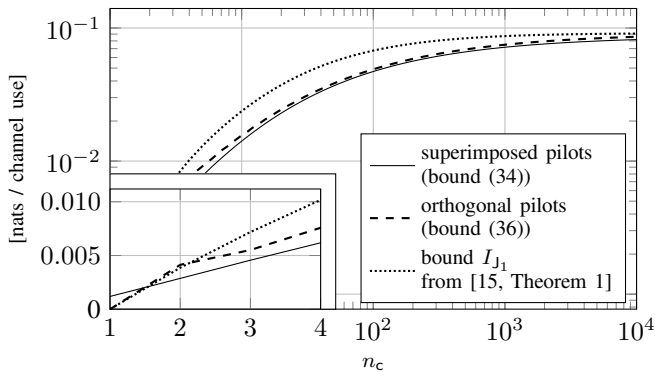
\begin{figure}[ht]
	\centering
	\begin{tikzpicture}
\begin{loglogaxis}[
	grid=major,
	height={4cm}, width={7cm},
	xlabel={$\nc$},
	ylabel={[nats / channel use]},
	enlarge x limits=false,
	xticklabels={,,$10^2$,$10^3$,$10^4$},
	yticklabels={,,$10^{-2}$,$10^{-1}$},
	legend style={cells={anchor=west, align=left}, fill=white, at={(0.98,0.02)}, anchor=south east, font=\footnotesize},
	label style={font=\footnotesize},
	scale only axis,
]
	\pgfplotstableread{coord/coord_RBF.txt}\mytable
	\addplot[solid] table[x=nc,y=I] {\mytable};
	\addplot[dashed,thick] table[x=nc,y=I_M_opt] {\mytable};
	\addplot[densely dotted,thick] table[x=nc,y=I_J] {\mytable};
	\addlegendentry{superimposed pilots\\(bound~\eqref{block_fading_lower_bound})}
	\addlegendentry{orthogonal pilots\\(bound~\eqref{block_fading_lower_bound_Medard})}
	\addlegendentry{bound $I_{\mathsf{J}_1}$\\from~\cite[Theorem~1]{MaJiLo09}}
	\draw[fill=white] (rel axis cs:0,0) rectangle +(3cm,1.8cm);
\end{loglogaxis}
\begin{axis}[
	grid=major,
	xmin=1, xmax=4, ymin=0,
	height={1.6cm}, width={2.8cm},
	yticklabel style={/pgf/number format/.cd,fixed,fixed zerofill,precision=3,/tikz/.cd},
	yticklabel={\ifdim\tick pt=0pt 0 \else\pgfmathprintnumber{\tick}\fi},
	scale only axis,
	scaled y ticks=false,
	restrict x to domain={1:5},
]
	\pgfplotstableread{coord/coord_RBF.txt}\mytable
	\addplot[dashed,thick] table[x=nc,y=I_M_opt] {\mytable};
	\addplot[densely dotted,thick] table[x=nc,y=I_J] {\mytable};
	\addplot[solid] table[x=nc,y=I] {\mytable};
\end{axis}
\end{tikzpicture}
	\vspace{-5mm}
	\caption{Comparison of achievable rates for orthogonal pilots vs.~superimposed pilots on a Rayleigh block-fading channel at an SNR of \SI{-10}{\decibel}. The plot region for low $\nc$ is zoomed in for better visibility. Note the double-logarithmic axes for the large-scale plot.}
	\label{fig:RBF_plot}
\end{figure}

Consider the single-antenna Rayleigh fading channel, which obeys the same system equation as~\eqref{system_equation} except that the fading gain $H$ is constant during $\nc$ consecutive time instants. The number $\nc$ is referred to as the {\em coherence} time.
By treating this block-fading channel as a $\nc \times \nc$ MIMO channel and particularizing Theorem~\ref{thm:MI_lower_bound_MIMO} in a suitable way, we obtain the following capacity lower bound based on superimposed pilots:
\begin{multline}   \label{block_fading_lower_bound}
	\frac{1}{\nc} I(\Bx;\By)
	\geq \frac{\nc-1}{\nc} \log\left( \frac{B}{B - \nc^2 \nu(1-\nu) \rho^2} \right) + \\
	\frac{1}{\nc} \log\left( \frac{A \nc + B}{A \nc + B - \nc^2 \nu(1-\nu) \rho^2} \right)
\end{multline}
where
\begin{subequations}
\begin{IEEEeqnarray*}{rCl}
	A
	&=& \bigl( 2\bar{\nu}^2 + 2\nc\nu(1+\nc\nu) + 2(2\nc-1)\nu\bar{\nu} - 3\nc\nu^2 \bigr) \rho^2 \\
	\IEEEeqnarraymulticol{3}{r}{
		{} + \nc\nu\rho
	} \IEEEeqnarraynumspace\IEEEyesnumber \\
	B
	&=& ( 2\nc(\nu\nc + \bar{\nu}) - \bar{\nu} ) \bar{\nu} \rho^2 + (2\bar{\nu} + \nu\nc) \nc\rho + \nc   \IEEEeqnarraynumspace\IEEEyesnumber
\end{IEEEeqnarray*}
\end{subequations}
We can read off~\eqref{block_fading_lower_bound} that for $\nc=1$ we recover the SISO bound~\eqref{MI_lower_bound} once we know that $\var(|Y|^2) = A+B$. The limit for $\nc \to \infty$ of the right-hand side of~\eqref{block_fading_lower_bound} is $\log\left( \frac{2(1-\nu) \rho + 1}{(1-\nu) \rho + 1} \right)$ which is maximized by setting $\nu=0$ (i.e., a vanishing pilot as $\nc \to \infty$) and gives $\log(2) = 0.6931$ nats, or $1$ bit, in the high-SNR limit. This clearly also highlights the limitation of superimposed pilots: for large $\nc$, our bound is markedly inferior to the time-multiplexed pilot bound~\cite{HaHo03}, \cite[Eq.~(5)]{MaJiLo09}
\begin{equation}   \label{block_fading_lower_bound_Medard}
	\frac{1}{\nc} I(\Bx;\By)
	\geq \max_{\tau \in \{1,\dotsc,\nc-1\}} \frac{\nc-\tau}{\nc} C\left( \frac{\rho^2 \tau}{1 + \rho(\tau+1)} \right)
\end{equation}
where $C(x) = \int_0^\infty \log(1+xt) e^{-t} \intd t$, whose limit as $\nc \to \infty$ is unbounded in the SNR $\rho$. This observation suggests that superimposed pilots tend to be competitive against orthogonal pilots mostly for very low channel coherence (approaching fast fading) and low SNR, which echoes similar conclusions known in the literature \cite{HoTu99,DoToSa04} and is evidenced by the plot in Figure~\ref{fig:RBF_plot}.

%

In this Figure, we compare the performance of superimposed pilots (bound~\eqref{block_fading_lower_bound} with optimized $\nu$) against time-multiplexed orthogonal pilots (bound~\eqref{block_fading_lower_bound_Medard}). We have further included a bound from~\cite{MaJiLo09} based on a single pilot symbol and so-called {\em joint pilot-data processing}, which implicitly reflects the gain that one could as well obtain by improving~\eqref{block_fading_lower_bound_Medard} via sequential reuse of decoded symbols as effective pilots, so as to gradually improve the channel estimate. We see that superimposed pilots are superior only for $\nc=1$ while achieving a rate close but inferior to that of orthogonal pilots elsewhere (although the gap widens as the SNR increases).

\section{Conclusion}

We have derived a new bound on the capacity for communication on noncoherent channels with superimposed pilots. Due to its simplicity, we have shown that it lends itself well to analytical treatment, for example, through closed-form optimization of the fraction of transmit power allocated to pilots. Its derivation is reminiscent of M\'edard's worst-case noise bound~\cite{Me00}, yet it differs in some crucial ways. We have further demonstrated how to refine the bound and generalized it to the MIMO setting. Let it be mentioned that another refinement method based on the rate-splitting approach from~\cite{PaKoFo14} is also fruitful, though beyond the scope of this article. We expect that, as with M\'edard's bound, our bounding technique enjoys a great potential in terms of wide applicability and extensibility to numerous scenarios of interest, such as semiblind channel estimation~\cite{KhBo07}, stationary fading with a power spectral density~\cite{Aste17}, massive MIMO~\cite{UpVoVe17} or multiuser settings, to only mention a few.

\bibliographystyle{IEEEtran}
\bibliography{IEEEabrv,custom/references}


\end{document}